\documentstyle[12pt,epsfig,epsf]{article}

\newcommand{\ba}{\begin{array}}
\newcommand{\ea}{\end{array}}
\newcommand{\bd}{\begin{displaymath}}
\newcommand{\ed}{\end{displaymath}}
\newcommand{\be}{\begin{equation}}
\newcommand{\ee}{\end{equation}}
\newcommand{\bea}{\begin{eqnarray}}
\newcommand{\eea}{\end{eqnarray}}

\def\gev{{\rm GeV }}
 
\def\db{{$\delta b$ }}

\def\q2 {q^2}

\begin{document}
\begin{center}
{\Large\bf Atmospheric Neutrinos as a Probe of CPT Violation}
\\[15mm]
\vskip 15pt
{\sf Anindya Datta $^{a, \,\!\!}$
\footnote{E-mail address: datta@roma1.infn.it }}, 
{\sf Raj Gandhi $^{b, \,\!\!}$
\footnote{E-mail address: raj@mri.ernet.in }}, 
{\sf Poonam Mehta $^{c,b \,\!\!}$
\footnote{E-mail address: mpoonam@mri.ernet.in }}, 
and 
{\sf S Uma Sankar $^{d, \,\!\!}$
\footnote{E-mail address: uma@phy.iitb.ac.in}}
\vskip 8pt
$^a${\em 
INFN, Sezione di Roma, 
Dip.~di Fisica, Universit\`a La Sapienza,
I-00185 Rome, Italy
}\\

$^b${\em Harish-Chandra Research Institute, Chhatnag Road, Jhunsi,
Allahabad 211 019, India
}\\

$^c${\em Department of Physics and Astrophysics, University of Delhi, 
Delhi 110 019, India
}\\

$^d${\em 
Department of Physics, I.~I.~T., Powai, 
Mumbai 400 076, India
}\\

\end{center}
\vskip 5pt 
\begin{abstract}

  We show that atmospheric neutrinos
can provide a sensitive and 
robust probe of CPT violation (CPTV). 
We perform realistic 
event-rate calculations and study the variations 
of the ratio of total muon to antimuon survival 
rates with $L/E$ and $L$ ($L$ $\equiv$ baseline length, $E$ 
$\equiv$ neutrino energy) in a detector capable of 
identifying the muon charge. 
We demonstrate that 
measurements of these ratios 
when coupled with the significant $L$ and $E$ range 
which characterizes the atmospheric neutrino spectrum provides 
a method of both detecting the presence of such violations and 
putting bounds on them which 
compare very favourably with those possible from 
a future neutrino factory.\\
\vskip 2pt
PACS numbers : 11.30.Er.,11.30.Cp.,14.60.Pq,13.15.+g
\end{abstract}
\vskip 0.5 cm
\setcounter{footnote}{0}

\begin{section}{Introduction}          
The CPT theorem is a cornerstone of quantum field theory 
in general and particle physics in particular. 
It rests on principles whose generality and scope makes 
them pillars of modern physics, like Lorentz invariance, 
the spin-statistics theorem, and the local and 
hermitian nature of the Lagrangian \cite{bjor}. 
Tests of CPT invariance thus assume importance 
not only because of the almost sacrosanct nature of these 
principles, but because any violation of CPT 
would signal radical new physics and force a 
re-thinking of foundational aspects of field theory 
and particle physics \cite{green,note1}. 
In particular, Greenberg \cite{green} has shown that
CPT violation necesssarily implies violation of Lorentz
invariance.

For over three decades, particle physics has focussed 
its efforts on testing the predictions of the 
Standard Model (SM) and seeking the next physics 
frontier beyond it. 
One of the conclusions to emerge from this 
multi-pronged effort over the past decade is that the 
neutrino sector, both via theory and via experiment, 
provides us with an almost unmatched window to physics 
beyond the SM \cite{barg3}.

The role  of neutrinos as probes of Lorentz and 
CPT invariance was discussed in a general framework 
by Colladay and Kostelecky \cite{coll} and by Coleman 
and Glashow \cite{cole}. Recent papers 
\cite{7to11}
have studied possible 
mechanisms beyond the SM which could lead to CPTV 
in the neutrino sector. CPTV, in the form of different 
masses for neutrinos and anti-neutrinos has been invoked 
to explain all the neutrino anomalies 
simultaneously, 
including the LSND result \cite{mur}. 
Bounds possible on such violations from reactor and 
solar experiments were discussed in \cite{bahcall} 
and the Dirac and Majorana nature of neutrinos in 
their presence was studied in \cite{kayser}.

As first proposed in \cite{barg} and also discussed in, for
instance, \cite{pak}, significant
bounds on CPTV parameters can be set in neutrino factory 
experiments due to their expected high luminosities and 
low backgrounds \cite{nufac}. However, with all their 
advantages, neutrino factories are a tool which may become 
available to us only about fifteen or twenty years 
from now.   

In contrast to this, detectors capable of accurately 
detecting the charge, direction and energy of a muon 
employ well understood and familiar technology. For instance, 
large mass magnetized iron calorimeter neutrino detectors were 
considered in \cite{mono} to study atmospheric neutrino 
interactions in great detail. At least one such detector is being 
currently actively planned to begin data-taking five years from now
{\cite{ino}}. 
We show that such detectors, or variants thereof, can, in 
conjunction with the by now well understood atmospheric 
neutrinos, form an ideal tool to detect CPTV in the neutrino sector. 
We focus on the survival probabilities for $\nu_\mu$ and
$\bar{\nu}_\mu$. A difference in these quantities is a 
signal for CPTV. By calculating the ratio of 
their event-rates, we show that comprehensive tests of 
CPTV are possible in the atmospheric 
neutrino sector, with sensitivities which compare very 
favourably with those projected for neutrino factory 
experiments. 
\end{section}

\section{CPT Violation in $\nu$ Interactions}

We consider the effective {\bf C and CPT-odd interaction} terms 
${{\bar{\nu}_L^\alpha  
b_{\alpha \beta}^\mu \gamma_\mu \nu_L^{\beta}}}$, 
where $\alpha$ and $\beta$ are flavour indices \cite{barg}. 
In presence of this CPTV term, the neutrino energy 
acquires an additional term which comes from the matrix $b^0_{\alpha \beta}$. 
For anti-neutrinos, this term has the opposite sign. 
The energy eigenvalues of neutrinos (in ultra-relativistic limit) are obtained by diagonalizing 
the Hermitian matrix given by
\begin{eqnarray}
{\rm {A}} &=& {\rm {\frac{m^2}{2p} + b}},
\end{eqnarray}
\noindent 
where ${{m^2 \equiv m\,m^\dagger}}$ is the Hermitian mass squared 
matrix and we have dropped the superscript $0$ from $b^0$.  

We assume equal masses for neutrinos and anti-neutrinos.  
For simplicity we have assumed that the two mixing angles 
that diagonalize the matrices ${{m^2}}$ and $b$ are equal 
(i.e. ${{\theta_m=\theta_b=\theta}}$).  
In addition, the additional phase that arises due to the two different 
unitary matrices needed to diagonalize the $\delta m^2$ 
and $\delta$b matrices\footnote{Only one of the two 
phases can be absorbed by a redefinition of neutrino states.} 
is set to zero.

For atmospheric neutrinos, it is at times 
({but not always}) a good approximation to consider 
two flavours only, depending on the parameters 
which one is studying. 
We adopt this in our calculations.  
This is tantamount to assuming that 
$\sin^2 2\theta_{13}$ is small ({below the CHOOZ \cite{chooz} 
bound}) and so is ${{\delta m^2_{21}}}$ 
({compared to ${{\delta m^2_{32}}}$}) 
and thus matter and related three-flavour 
effects can be safely neglected. 
As shown in \cite{bern} matter effects show up in 
atmospheric neutrinos for $\sin^2 2\theta_{13} \sim 0.1$ 
and baselines above 7000 km. 
The expression for survival probability 
for the case of CPTV 2-flavour 
oscillations then becomes
\begin{eqnarray}
{\rm P}_{\alpha \alpha}({\rm L}) = 
{\rm {1 - \sin^2 2\theta}} 
~{\rm {\sin^2 \left
[\left( \frac{\delta m^2}{4 E} + 
\frac{\delta b}{2} 
     \right) L \right]}}
\label{prob}
\end{eqnarray}
where ${{\delta m^2}}$ and \db 
are the differences between the eigenvalues of the 
matrices ${{m^2}}$ and $b$, respectively and 
$\alpha$ corresponds to $\mu$ or $\tau$ flavours. 
Note that \db has units of energy (GeV).  
For $\bar\nu$, the sign of \db is reversed. 
The difference between 
${\rm P}_{\alpha \alpha}$ and 
${\rm P}_{\bar\alpha \bar\alpha}$
is given by,
\begin{eqnarray}
\Delta {\rm P}_{\alpha \alpha}^{\rm CPT} 
&=& 
- {\rm {\sin^2 2\theta \,\sin \left(
\frac{\delta m^2 L}{2E} \right) 
\,\sin ( \delta b L )}} 
\end{eqnarray}


An important consequence of the modified dispersion relation 
in presence of CPTV is that 
the characteristic $L/E$ behaviour of neutrino oscillations is lost. 
Hence depending on which term is larger for a given set of parameters 
and the energy, the mixing angle and oscillation length can vary 
dramatically with E. Thus precision oscillation measurements can set 
unprecedented bounds on such effects. 
Also, in order 
to see any observable effect of CPTV, one must have both CPT-even and 
CPT-odd terms to be non-zero. 
 
\section{Calculations}
In order to quantitatively demonstrate the feasibility 
of using atmospheric neutrinos as a source of detecting 
and putting bounds on CPT violation, we 
focus on a typical detector which can detect muon energy 
and direction and also identify its charge. The simplest 
choice of a suitable prototype is an iron calorimeter, 
which employs well-understood technology. Such a detector 
was proposed for  Gran Sasso (MONOLITH) \cite{mono} and 
is also currently being planned for a location in India 
(INO) \cite{ino}, with initial data-taking by 2007. 
It is contemplated as both a detector for atmospheric 
neutrinos and as a future end detector for a neutrino 
factory beam.

The atmospheric neutrino physics program 
previously studied in the literature in the context of 
a Magnetized Iron Tracking Calorimeter 
includes attempting to obtain conclusive proof
that neutrinos oscillate by 
observation of a $L/E$ dip in the up-down
ratio of atmospheric neutrino induced muons, and a more 
accurate pinning down of oscillation parameters.
However, its usefulness as a detector for CPTV parameters using atmospheric neutrinos has not 
been studied earlier. 

Our prototype is a 50 kT Iron 
detector, with detection and charge discrimination 
capability for muons, provided by a B field of about 
1.2 Tesla. We have assumed a 
(modest) 50\% efficiency of the detector for 
muon detection and a muon energy resolution of within 5\%. 
We have factored in a resolution in $L/E$ of 
50\% at Full Width Half Maximum, and incorporated the 
requisite smearing in our event-rate calculations. 
The resolution function is best parameterised 
by an exponential damping term given by, 
${ {R(\delta m^2, L/E ) }} = \exp({-0.25\, {{\delta m^2 L/E}}})  $ 
\cite{mono}.

In the calculations presented here, we have assumed that 
the atmospheric neutrino problem is resolved by 
$\nu_{\mu} \rightarrow \nu_{\tau}$ oscillations.
Specifically, we use the following input parameters : 
${{\delta m^2_{32} = 0.002 ~{\rm {eV}}^2}}$, 
$\sin^2 2\theta_{23}$ = 1, which are consistent 
with best fit values determined by the most recent analyses of 
atmospheric data combined with CHOOZ bounds \cite{lisi}. 
In addition we have used the Bartol atmospheric flux 
\cite{bartol} and set a muon detection threshold of $1$ GeV. 
For neutrino energies below $1.8$ GeV the 
quasi-elastic $\nu$-nucleon crosssection has been 
used, while above this energy we have put in the 
DIS value of the crosssection. 
The number of muon events have been calculated using 

\begin{eqnarray}
N = N_n \times M_d \int \sigma_{\nu_\mu - N}^{CC} \,P(\nu_\mu \to \nu_\mu) \,
\frac{dN_\nu}{dE_{\nu}}\, dE_{\nu} 
\end{eqnarray}
\noindent where $N_n=6.023\times 10^{32}$ is the number of (isoscalar) nucleons in 1kT of target material and $M_d$ is the detector mass. Our results are obtained from a simple parton level monte-carlo event generator. We have used CTEQ4LQ 
\cite{lai}
parametrisations for the parton distribution functions to estimate the DIS crosssection.

Finally, we comment on the exposure time necessary to 
see a dependable signal. Since the number of 
$\bar\nu$ atmospheric events will be significantly 
smaller than the number of $\nu$ 
events, reducing the statistical error in the ratio 
will require an exposure time that enables observations of a 
sufficient number 
of $\bar \nu$ events. 
Our calculations indicate that an exposure 
of $400$ kT-yr would be sufficient for 
statistically significant signals to emerge. 

\vskip .4in
\begin{figure}[h]
\includegraphics[scale=.65]{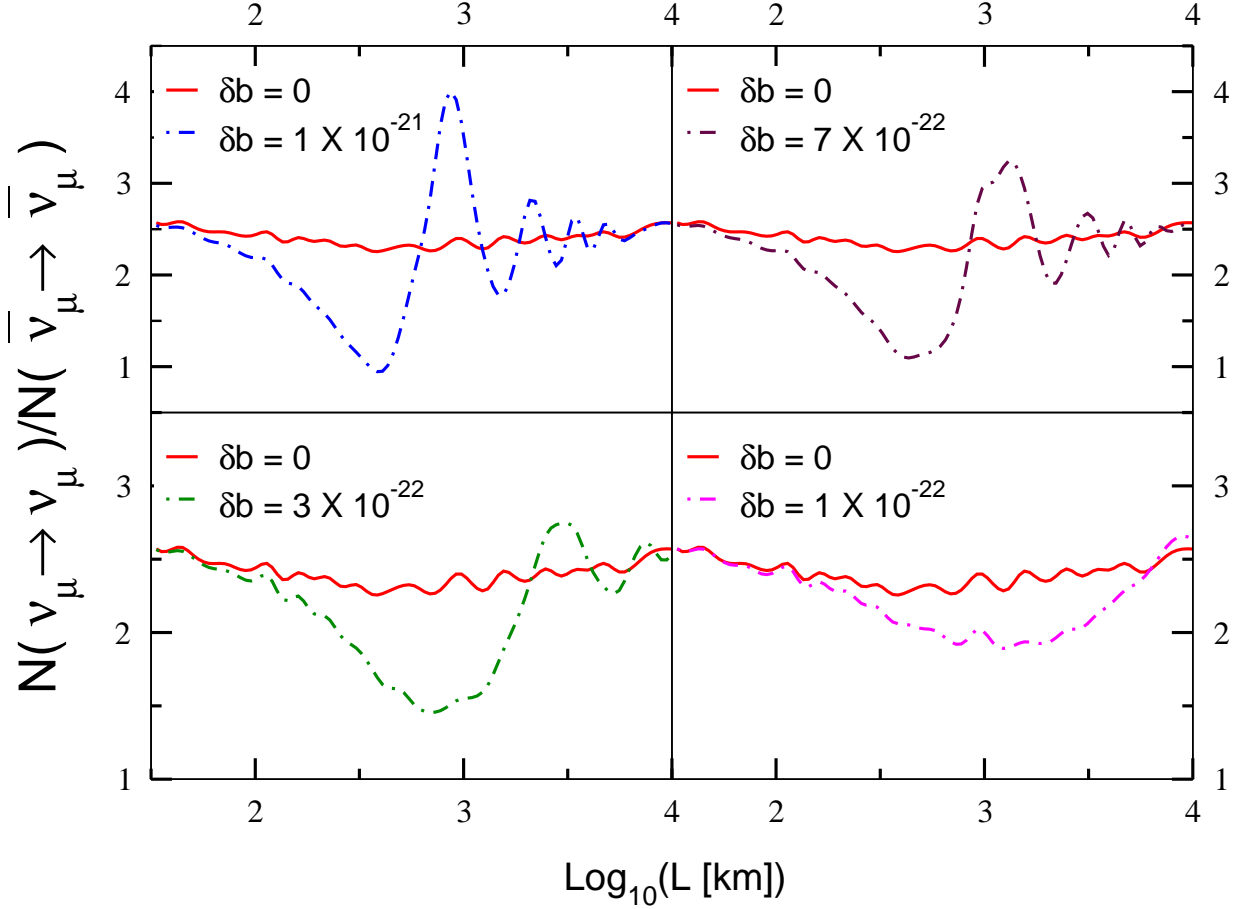}
\caption{\label{fig:fig1} {\em {The ratio of total 
muon to anti-muon events plotted against ${Log_{10}(L)}$ 
for different values of $\delta b$ (in GeV). The oscillation parameters used in all the plots are : $\delta m^2 = 2\times 10^{-3}$ eV$^2$ and $\sin^2{2\theta_{13}}=1$.
}}}
\end{figure}

\begin{figure}[h]
\includegraphics[scale=.65]{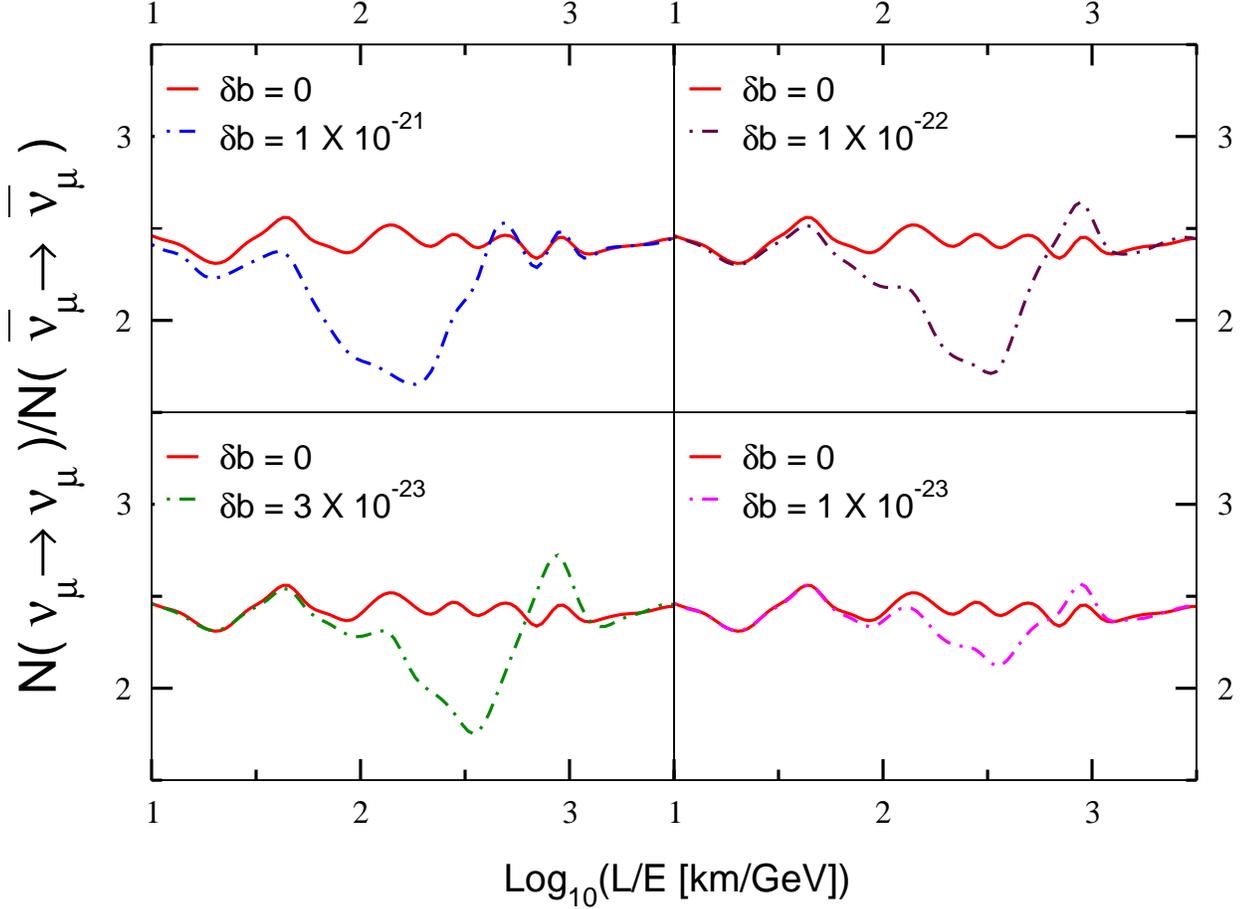}
\caption{\label{fig:fig2} 
{\em {The ratio of total 
muon to anti-muon events plotted against ${Log_{10}(L/E)}$ for 
different values of $\delta b$ (in GeV).
}}}
\end{figure}

\section{Results and Discussions}

Figure~\ref{fig:fig1} shows the variation of the ratio of 
total (up + down) muon survival events to those of anti-muons, 
plotted vs $L$ for various values of $\delta b$. 
The solid line in each of the plots is the (CPT conserving) 
\db= 0 case, shown for comparison. 
The overall shape and position of this (solid) curve is 
representative of the ratio of the two crosssections 
($\nu$ vs $\bar\nu$) at the relatively low (few GeV) 
energies which dominate the event-rates.
The small wiggles and variations are a result of the various 
energies and lengths involved and angular differences in 
fluxes which characterize the overall atmospheric neutrino 
spectrum.


From Equation~3, we see that the CPTV 
difference in probabilities will become zero 
whenever ${{\delta b L = n \pi}}$ ($n$=integer), 
resulting in a node ({\it i.e.} an
intersection with the ${{\delta b=0}}$ curve).  
The positions and the
number of nodes for the various curves nicely 
correspond to these
expected ``zeros'' of CPTV and also provide a way of
distinguishing between them. 
Clearly, parameter values of the order of 
${ {\delta b = 3 \times 10^{-22}}}$ \gev 
should be nicely discernible in these observations. 
We note that here  the effects of the CPTV 
parameters are maximal at baseline length 
$L\approx 1000$ km, thus neglecting matter 
effects is justified even if $\theta_{13}$ is close to the 
CHOOZ upper bound. 

In Figure~\ref{fig:fig2}, we plot the same ratio of event-rates vs
$L/E$. The nodal position is now dictated by the term ${{\sin (\delta
m^2 L/2E)}}$, resulting in a common node for the various $\delta b$
values at ${{\delta m^2 L/2E=n\pi}}$. The plots also show a
significant dip near ${{L/E\simeq310}}$ km/GeV. This is explained by
the fact that ${{\delta m^2 L/ 4 E = \pi/4}}$ for this value. In
Equation~2, the sine function has its maximum slope at this value of
its argument, and hence the survival probabilities for $\nu$ and
$\bar\nu$ differ maximally here due to the sign difference of the \db
terms, providing highest sensitivity to the presence of CPTV
parameters. We note that in the vicinity of the dip the antineutrino
event-rate increases and the neutrino rate decreases, which
consequently tends to reduce the statistical error in the ratio,
aiding detection. This set of curves provides heightened sensitivity
to the ${\it presence}$ of CPTV, without the same discriminating
sensitivity (between various \db values) of the plots in Figure~1. For
instance, CPTV induced by parameter values as low as ${{\delta b = 3
\times 10^{-23}}}$ can be detected.  For a lower value of \db, say
$10^{-23}$, the curve tends to creep back closer to the ${{\delta b =
0}}$ solid line. Very recently, the authors of ref. \cite{maltoni} considered
the bounds on various types of new physics coming from Super-K and K2K
data.  They obtain $\delta b \leq 5 \times 10^{-23} {\rm ~GeV}.$



{\bf Conclusions :} Atmospheric neutrinos in a detector 
capable of measuring muon energy and direction and identifying 
its charge can allow us to set significant bounds on 
all types of CPTV in the neutrino sector. 
These bounds compare very favourably with those 
possible from future neutrino factories \cite{barg}. 
Specifically, the charge discrimination capability 
of such a detector when coupled with the significant $L$ and 
$E$ ranges which characterize the atmospheric neutrino 
spectrum provides a potent and sensitive probe 
of such violations. 
By calculating the ratios of muon 
and anti-muon events and studying their variation 
with $L$ and $L/E$ we have shown that the presence of 
CPTV can be detected provided 
${{\delta b > 3 \times 10^{-23}}}$ \gev. 
For somewhat higher values of $\delta b$,  
it is also possible to obtain a measure of their 
magnitudes by studying their minima and zeros as 
discussed in the text. 

\vskip 0.1in
{\bf Acknowledgments :} PM would like to acknowledge 
Council for Scientific and Industrial Research, India for 
partial financial support and thank 
HRI and INO for hospitality. 

\end{document}